\documentclass[twocolumn,showpacs,amsmath,amssymb,prl,superscriptaddress]{revtex4}
\usepackage{graphicx}
\usepackage{bm}
\usepackage{dcolumn}

\begin{document}
\title{Dual nature of the ferroelectric and metallic state in LiOsO$_3$}

\author{Gianluca Giovannetti}
\affiliation{CNR-IOM-Democritos National Simulation Centre and International School for Advanced Studies (SISSA), Via Bonomea 265, I-34136, Trieste, Italy}
\affiliation{Institute for Theoretical Solid State Physics, IFW-Dresden, PF 270116, 01171 Dresden, Germany}
\author{Massimo Capone}
\affiliation{CNR-IOM-Democritos National Simulation Centre and International School for Advanced Studies (SISSA), Via Bonomea 265, I-34136, Trieste, Italy}

\begin{abstract}
Using density functional theory we investigate the lattice instability and electronic structure of recently discovered ferroelectric metal LiOsO$_3$. We show that the ferroelectric-like lattice instability is related to the Li-O distortion modes while the Os-O displacements change the d-p hybridization as in common ferroelectric insulators.
Within the manifold of the d-orbitals, a dual behavior emerges. The ferroelectric transition is indeed mainly associated to the nominally empty e$_g$ orbitals which  are hybridized with the oxygen p orbitals, while the t$_{2g}$ orbitals are responsible of the metallic response. Interestingly, these orbitals are nominally half-filled by three electrons, a configuration which suffers from strong correlation effects even for moderate values of the screened Coulomb interaction.
\end{abstract}

\pacs{}
\maketitle

Ferroelectric and multiferroic materials represent a fascinating ground where an immense applicative potential\cite{Cheong} emerges from the competition between different interactions with an intimate interplay of spin, charge, orbital, and lattice degrees of freedom. 
In a prototype example of ferroelectric system, the d$^{0}$ perovskite BaTiO$_3$, the ferroelectic distortion is driven by the  hybridization between the filled oxygen 2p states and the nearly empty d states of the transition metal (TM) cation. The ferroelectric displacement reduces the distance between the TM cation and one or more of the surrounding oxygen anions leading to gain in covalent bond energy. This effect leads to a maximum energy gain for empty d orbitals, because the oxygen electrons populate only the bonding combination, but it is expected to survive for sufficiently low filling of the TM orbitals, when only a small fraction of the antibonding combinations is populated. The fact that this mechanism for ferroelectricity requires a nearly unoccupied d shell of the TM cations while a partial filling of these electronic d states is required to have a  magnetic moment  explains the incompatibility between ferroelectricity and magnetism and the reason of scarcity of multiferroic materials. 

From this perspective, CaMnO$_3$ is an interesting counterexample. Here Mn is an a $d^3$ configuration yet it simultaneously displays ferroelectricity and magnetic ordering. This is understood by observing that the empty e$_g$ orbitals hybridize with the oxygen p orbitals leading to ferroelectric distortions, while the half-filled t$_{2g}$ orbitals give rise to the magnetic ordering\cite{Khomskii,Ederer}. Therefore, while BaTiO$_3$ is a band insulator due to the  d$^{0}$ configuration, the insulating state of CaMnO$_3$ is dominated by the strong correlation in the $t_{2g}$ manifold\cite{Luo}.

On the other hand, metals are not expected to exhibit ferroelectricity because the itinerant electrons screen the electric fields and inhibit the electrostatic forces responsible for ferroelectric distortions. The concept of a  "ferroelectric metal" was however theoretically proposed in 1965 by Anderson and Blount 1965 \cite{Anderson}, who have shown that the loss of a center of symmetry in a metal is possible only through a second-order phase transition.
The pyrochlore compound Cd$_2$Re$_2$O$_7$ appeared as possible realization because it displays an unusual structural transition in its metallic state, but a unique polar axis could not be identified \cite{Sergienko,Tachibana}. For electron-doped BaTiO$_3$ ferroelectric distortions occur in two distinct metallic and insulating phases that do not coexist microscopically \cite{Kolodiazhnyi,Jeong}. 

A first success has been recently obtained in the search for a ferroelectric metal. LiOsO$_3$ displays indeed metallic conduction while a second-order phase transition to a state where the ionic structure breaks the lattice symmetry occurs at T$_s$=140K\cite{Shi}.  Neutron and x-ray diffraction studies show that the space group changes from centrosymmetric R$\overline{3}$c to ferroelectric-like R${3}$c \cite{Shi}. 

\begin{figure}
\includegraphics[width=.99\columnwidth,angle=-0]{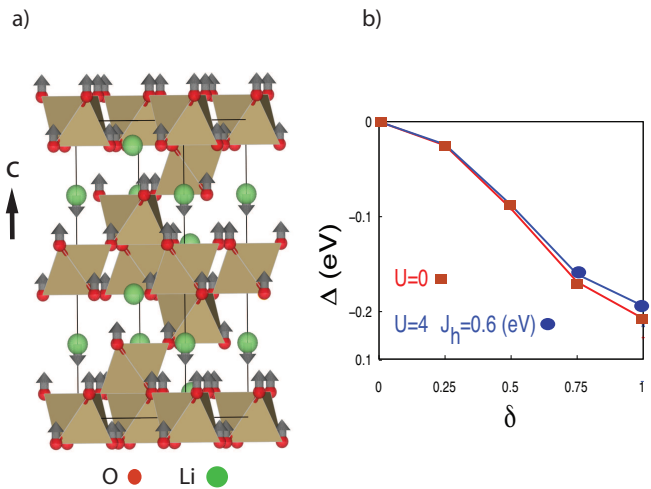}
\caption{(Color online) (a) Ionic displacements (arrows) along the polar axis $c$ in the non centrosymmetric structure of LiOsO$_3$ (b) Total energy gain ($\Delta$) due to the ferroelectric-like distortions calculated within LDA as a function of the distortion $\delta$ ($\delta=1$ corresponds to the experimental distorted structure.)}
\label{fig1}
\end{figure}

The crystal structure of LiOsO$_3$ consists of oxygen octahedra sharing edges (see Fig. \ref{fig1}a).
In the R$\overline{3}$c centrosymmetric structure the Os atoms are at the center of the octahedra having all equivalent distances with O sites, while Li atoms are centered between two adjacent oxygen octahedra along the c axis. The loss of symmetry involves the ferroelectric displacements of Li and O atoms along the polar axis $c$: the Os atom gets off-centered and Li atom is ferroelectrically displaced with respect to the oxygen planes (see Fig. \ref{fig1}a)). This structural transition is very similar to that observed in the isostructural compound LiNbO$_3$, where the ferroelectric state is understood in terms of Nb-O hybridization\cite{Inbar}.

Here we address the driving force behind the ferroelectric instability and we study the electronic structure of metallic LiOsO$_3$. We find the ferroelectric-like distortions to be mainly related to the Li-O modes while Os-O displacements enhance the hybridization between the $e_g$ orbitals of Os and the p-orbitals of O.  The three electrons populating the $t_{2g}$ orbitals are instead responsible for the metallic behavior. As opposed to case of CaMnO$_3$, the strength of the electron-electron interactions is not sufficient to drive a metal-insulator transition, and the system remains a correlated metal. Interestingly, such correlated state is not unstable to a Stoner-like antiferromagnetic state because the bandstructure strongly breaks particle-hole symmetry. The stabilization of the ferroelectric structure in LiOsO$_3$ is the consequence of the itinerant electrons nearly decoupled from the soft phonons which break the inversion symmetry.

We start our investigation performing density functional theory calculations (DFT) \cite{DFT} for the refined experimental structures with space groups R$\overline{3}$c and R3c using the Vienna \textit{ab-initio} simulation package\cite{VASP}. 
Our DFT simulations are performed within the local density approximation (LDA)\cite{lda} to the exchange-correlation potential. In a second step we supplement the LDA calculations by including also a  Hubbard-like interaction and the Hund's exchange J$_H$ on Os d-orbitals within the LDA+U scheme\cite{ldaplusU}. The cut-off for the plane-wave basis set is 500 eV and a $12 \times 12 \times 8$ mesh is used for the Brillouin-zone sampling. We perform atomic relaxations until until the change in total energy is less than 10$^{-6}$ eV. The check of the symmetry is performed with the program FINDSYM \cite{FINDSYM}.

To study the structural ferroelectric instability we consider the refined experimental structure with space group R${3}$c and we relax the ionic coordinates. Our optimized internal coordinates accurately recover the experimental R3c structure\cite{results_relaxation}, and provide results in agreement with previous ab-initio calculations showing that the soft A$_{2u}$ mode at the zone center is responsible for the non-centrosymmetric transition\cite{Kim,Xiang}. Constructing a putative structure with R$\overline{3}$c symmetry \cite{centrosymmetric_structure} using the program PSEUDO \cite{pseudo}, we identify that the ferroelectric-like distortions are of mirror-symmetry along the $a$ and $b$ unit cell vectors while along the polar axis $c$ the displacements of Li, Os and O ions are respectively 0.449, 0.002, -0.028 $\AA$. The breaking of the crystal symmetry is due to the different magnitude and direction of the Li and O displacements.

We define $\delta$ as a parameter measuring the distortions such that $\delta$=0 corresponds to the centrosymmetric ionic positions while $\delta$=1 corresponds to the ferroelectric ionic positions. The variation of the total energy $\Delta = E_(0) - E(\delta)$ (E($\delta$ ) being the total energy for a given distortion) along the path connecting the centrosymmetric ($\delta$=0) and ferroelectric ($\delta$=1) ionic structures show that the fully distorted structure is clearly energetically favored and it gains 0.2 eV per unit cell within LDA, as shown in Fig. 1(b)  (note that there are 6 Os ions in the unit cell).

\begin{figure}
\includegraphics[width=.825\columnwidth,angle=-90]{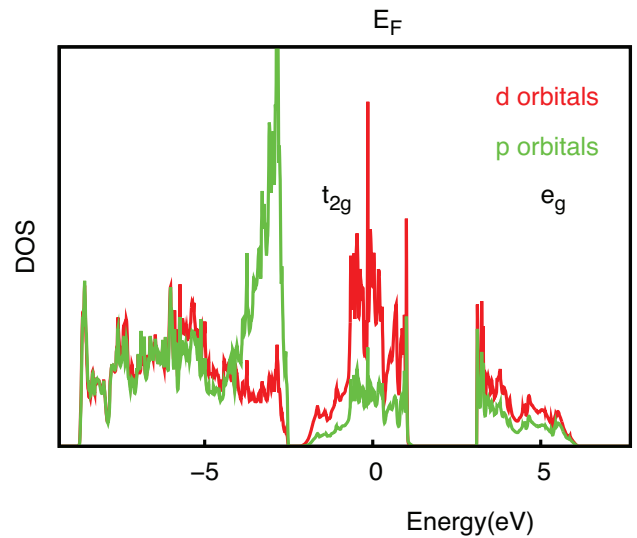}
\caption{(Color online) Orbital-resolved density of states for d and p orbitals within LDA. The Fermi level is the zero of the energy axis.}
\label{fig2}
\end{figure}

The density of state we obtain in LDA, shown in Fig. \ref{fig2} is clearly that of a metal, in agreement with the experiments and previous electronic-structure calculations \cite{Shi,Kim,Xiang}. The orbital-resolved density of states shows that there is large hybridization between the d orbitals of Os atoms and p states of O ions and this hybridization does not change much during the ferroelectric transition as it happens in other ferroelectrics BaTiO$_3$, PbTiO$_3$ and LiNbO$_3$ \cite{Inbar,Cohen}. 

We derive a tight-binding model by building maximally localized Wannier orbitals\cite{wannier90} choosing an energy window which includes bands originating from all the d orbitals of Os around the Fermi level (t$_{2g}$ and e$_g$ manifolds in Fig. \ref{fig2}). A first observation is that  the crystal-field splitting between the t$_{2g}$ and e$_g$ orbitals is rather large ( $\sim$ 4 eV) and gives rise to a clear separation between the spectral feature around the Fermi level and a higher-energy band.
Os has a nominal valence of 5+, so that 3 electrons have to fill the  t$_{2g}$ and e$_g$ orbitals.
In the energy window labeled as t$_{2g}$ around the Fermi level (see Fig. \ref{fig2}) 18 bands are present, resembling an half-filled configuration of a three-fold manifold. At higher energy we find the unoccupied bands with mainly e$_{g}$ character. 
It is important to notice that refer to t$_{2g}$ and e$_g$ orbitals, even in a case where, strictly speaking, the ferroelectric distortion mixes the two manifolds. Yet, the separation between a three-fold manifold and a two-fold one is evident in the bandstructure.
\begin{figure}
\includegraphics[width=.75\columnwidth,angle=-90]{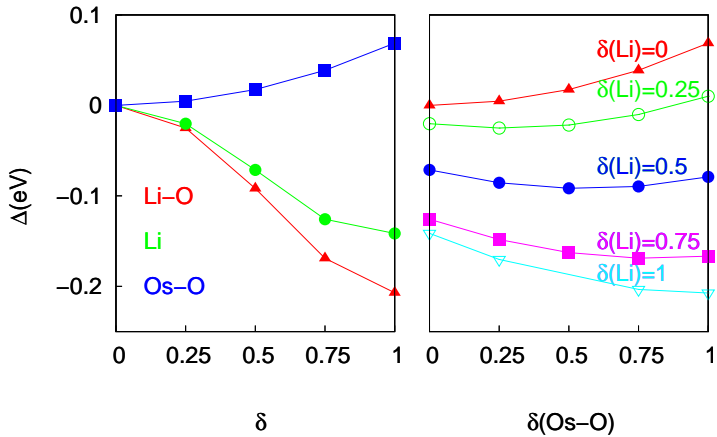}
\caption{(Color online) (a) Total energy gain $\Delta$ as a function of the displacements of selected atoms.. We compare the results obtained for Li-O, Os-O and Li modes, showing that most of the energetic gain comes from the Li-O mode. (b) Total energy gain  as a function of the displacement of the Os-O mode ($\delta(Os-O)$) for different fixed value of the Li-O displacement ($\delta(Li-O)$). The Os-O displacement leads to an energy gain only if the Li atoms is displaced.} 
\label{fig3}
\end{figure}

The d$^{3}$ configuration of LiOsO$_3$ with "t$_{2g}$" close to half-filled and "e$_g$" close to be empty, as for CaMnO$_3$ \cite{Khomskii,Ederer}, can be described as a  e$_g^{0}$ electron configuration, with an empty set of bands, analogously to the d$^{0}$ configuration of most perovskite ferroelectric insulators. Then the empty e$_g$ states can in principle provide the increase in bond energy required for the ferroelectric instability by hybridizing with the p-orbital.  


In order to identify how the above described electronic structure leads to the ferroelectric instability, in Fig \ref{fig3}a) we show the total energy variations $\Delta$ as a function of the ferroelectric distortions separated in contributions coming from Li-O, Li, Os-O modes. Li-O mode refers to the case where the Os displacement is neglected, Li mode refers to the case where only Li displacements are considered,  while in the case of Os-O mode the Li displacements are neglected. The energy gain is dominated by the displacement of the Li and O ions with the energetic contribution of O ions being smaller with respect to Li ions. Indeed the pure Li-O mode accounts for almost the full energy gain due to the ferroelectric distortion of LiOsO$_3$. The Li and O displacements are then responsible for the ferroelectric transition in LiOsO$_3$ as already pointed from previous density functional theory calculations \cite{Kim,Xiang}.

On the other hand the Os-O mode leads to an energy increase in the absence of Li displacement. The situation changes, as shown in panel b) of Fig. \ref{fig3}, when the Li atom is displaced. Indeed, plotting $\Delta$ as a function of the ferroelectric distortions of the pure Os-O modes ($\delta$(Os-O)) for different fixed values of the Li displacement, we find that the Os-O mode leads to an energy gain when the Li atoms are also displaced, with a maximum gain for the fully displaced case ($\delta$(Li) =1). 
Our results demonstrate that there is a cooperative and coupled displacement of Li and O ions, such that the shift of Li atoms gives the possibility to gain in the energy from the d-p hybridization of Os and O atoms. The polar distortion is then also stabilized via enhanced Os-O hybridizations. The latter mechanism is common in many different ferroelectric insulators and in particular in LiNbO$_3$ \cite{Inbar,Cohen}.

\begin{figure}
\includegraphics[width=.99\columnwidth,angle=-0]{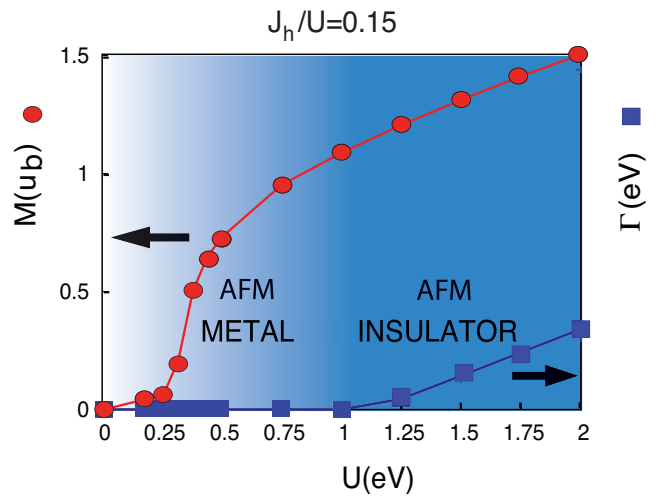}
\caption{(Color online) Magnetization M and charge gap $\Gamma$  as a function of U within LDA+U calculations at fixed ratio J$_H$/U =0.15.}
\label{fig4}
\end{figure}

Despite its metallic nature, LiOsO$_3$ has a residual resistivity more than one order of magnitude larger that that expected for a normal metal \cite{Shi} and the spin-susceptibility shows a Curie-Weiss behavior signaling the existence of local magnetic moment, even if an  ordered magnetic moment larger than 0.2 u$_B$ is experimentally excluded\cite{Shi}. This observation suggest a bad metallic behavior, which can be ascribed to the effects of electron-electron correlations.

Indeed, although the 5d orbitals are usually spatially extended and in LiOsO$_3$ and the delocalization is even more enhanced from hybridization with oxygen orbitals, electron correlations can still play an important role\cite{Shinaoka} in 5d systems, and in particular in LiOsO$_3$ with t$_{2g}$ orbitals close to half-filling. 
The accurate value of the Hubbard U is not known for perovskite osmates, but it is not expected to exceed the range of values estimated for the  iridates Sr$_2$IrO$_4$ and Ba$_2$IrO$_4$ \cite{irU}, respectively 1.4 and 2.4 eV. This value has to be compared with an LDA bandwidth of around 3eV for the t$_{2g}$ bands. We now investigate whether this intermediate level of correlation affects the electronic properties of LiOsO$_3$.

As a first step, we perform local spin-density approximation (LSDA) calculations with ferromagnetic (all spins are parallel) and G-type antiferromagnetic (every spin is antiparallel to all its neighbors) magnetic structures without finding any stable solution, ruling out a pure Slater antiferromagnetism.
We then perform calculations by including the local Coulomb interactions between the Os d electrons, namely the Hubbard U and the Hund's coupling starting from the experimental structure. $U$ ranges from 0 to 2 eV and the ratio $J_H/U$ is fixed at 0.15. We have checked that our conclusions are not strongly dependent on the value of  J$_H$. The spin-orbit coupling is not included, assuming that it has a limited effect as in the Slater insulator NaOsO$_3$\cite{Yongping}.

The energy gain due to the ferroelectric distortion is slightly reduced by electronic interactions (see Fig. \ref{fig1}b)). Within LSDA+U calculations a {\it metallic} G-type magnetic structure with local magnetic moments $M$ smaller that 1 u$_b$ can be stabilized at U values smaller than 1 eV (see Fig. \ref{fig4}). At a moderate value of U close to 1 eV, the G-type solution becomes insulating with a small but finite charge gap $\Gamma$ (see Fig. \ref{fig4}). As shown in Fig. \ref{fig4} M and $\Gamma$ both increase upon increasing the electron-electron interactions U and J$_H$. 

The Hartree-Fock treatment of correlations in LSDA+U naturally overestimates ordered phases, which have not been observed experimentally (even if an ordered magnetic moment smaller than 0.2 u$_B$ is compatible with the experimental statistics\cite{Shi}). On the other hand, these calculations show that the half-filled t$_{2g}$ manifold is sensitive to correlation effects, which appear as the underlying physics behind the bad metallic behavior with local magnetic moments.

Our calculations therefore highlight a significant role of electron-electron correlations, even if the strength is not sufficient to drive a Mott transition in the correlated manifold. From this point of view, the theoretical description of LiOsO$_3$ and of its metallic phase requires a more accurate treatment of correlations such as dynamical mean-field theory. In this delicate metallic regime, It would also be important to accurately estimate the values of the interaction terms $U$ and $J_H$ and the spin-orbit coupling. 

In conclusion, we investigate the ferroelectric transition and electronic structure of LiOsO$_3$. We find the Li-O distortion modes to be responsible for ferroelectric-like instability while the Os-O distortions allow for the hybridization of Os d states and O p states as in common ferroelectric insulators. The nearly empty e$_g$ orbitals hybridize with the oxygen p orbitals leading to the ferroelectric distortions, while the nearly half-filled t$_{2g}$ orbitals are associated to the metallic response.  Our study implies that the lattice and electronic degrees of freedom involved respectively in  the ferroelectric transition and metallic ground state are weakly coupled in the second-order ferroelectric-like phase transition. 
The metallic state is associated to a nearly half-filled manifold of "t$_{2g}$" bands, in which the moderate electron-electron interactions lead to noticeable correlation effects, which lead to a bad metallic behavior with relatively high resistivity and fingerprints of local moments, even in the absence of a sizable ordered magnetic moment. 

We acknowledge financial support by European Research Council under FP7/ERC Starting Independent Research Grant ``SUPERBAD" (Grant Agreement n. 240524). Calculations have been performed at CINECA (HPC project lsB06\_SUPMOT).


\begin{thebibliography}{99}
\bibitem{results_relaxation} Using the lattice vectors of the R3c non-centrosymmetric experimental crystal structure the optimized Wyckoff positions are: z$_{Li}$=0.88274, z$_{Os}$=0.66651 and x$_O$=0.29572, y$_O$=0.33011, z$_O$=0.25210. The experimental Wyckoff positions are: z$_{Li}$=0.88137, z$_{Os}$=0.66667 and x$_O$=0.29267, y$_O$=0.32313, z$_O$=0.25250.
\bibitem{centrosymmetric_structure} The Wyckoff positions of the putative centrosymmetric structure are: z$_{Li}$=0.91666, z$_{Os}$=0.66666 and x$_O$=0.29483, y$_O$=0.33333, z$_O$=0.25000.
\end{thebibliography}
\end{document}